\documentclass[11pt,twoside]{article}

\usepackage{asp2004}
\usepackage{epsf}
\usepackage{graphicx}
\usepackage{lscape}
\newcommand{\cd}{CD~$-24^\circ 731$}
\newcommand{\pg}{PG~1219$+$534}
\newcommand{\cpd}{CPD~$-64^\circ 481$}

\newcommand{\pb}{$^{208}$Pb}

\begin{document}
\title{Abundances of heavy metals and lead isotopic ratios in
   subluminous B stars}   
\author{S. O'Toole$^{1,2}$, U. Heber$^{1}$} 
\affil{$^1$Dr. Remeis-Sternwarte, Astronomical Institute, University of
	Erlangen-N\"urnberg, Sternwartstr. 7, D96049 Bamberg, Germany}  
\affil{$^2$Anglo-Australian Observatory, PO Box 296 Epping NSW 1710}   

\begin{abstract} 
  We present a detailed abundance analysis of high-resolution
  ultraviolet echelle spectra of five subdwarf B stars obtained with HST-STIS  
  The goal of our observations was to
  test the hypothesis that pulsations in sdBs are correlated to the
  surface abundances of iron-group elements. We study two
  pulsators and three non-pulsators and determined abundances for 
  25 elements including the iron group and 
  even  heavier elements such as tin and lead using LTE  
  spectrum 
  synthesis techniques. We find strong enrichments of heavy elements up to 
  2.9~dex with respect to solar which are probably caused by atomic diffusion 
  processes.
  No clear-cut correlation
  between pulsations and metal abundances
  becomes apparent 
  Abundances for lead isotopes are derived from very high resolution spectra 
  using an UV line of triply ionised lead. As Pb terminates the s-process sequence 
  Pb isotopic abundance ratios yield important constraints. It is very difficult 
  to measure them in hot stars. For the first time we were able to measure them 
  in two subluminous B stars and conclude that the $^{207}$Pb/$^{208}$Pb is
  solar.
\end{abstract}



\section{Introduction}


The possibility of pulsations in subluminous B stars was theoretically 
predicted by
Charpinet et al. (1996) at around the same time they were observed by
Kilkenny et al. (1997). The more than 30 known pulsators 
(V361\,Hya stars) have $T_{\mathrm{eff}}=29\,000-35\,000$\,K and
$\log g=5.2-6.0$, periods of 1-10 minutes and amplitudes less than 60\,mmag.
The richness of the pulsation
modes makes these stars ideal targets for asteroseismolgy.
 The driving mechanism of the
oscillations is believed to be related to the ionisation of iron and other
heavy elements at the base of the photosphere. As is the case
for other types of pulsators there is
an overlap in the ($T_{\mathrm{eff}}, \log g$) plane between pulsators and
non-pulsators. Diffusion calculations by Charpinet et al. (1997)
suggest that the surface iron abundance of pulsators should be higher than
that of non-pulsators, however studies by Edelmann et al. (2006) and
Heber et al. (2000) find that iron is approximately solar
in most sdBs.
For this reason we set out to determine if any correlation exists between
surface abundances of iron-group elements for pulsators and
non-pulsators. 
Since elements such as Ni, Mn and Cr are not
normally accessible through ground-based optical spectra, it was necessary to
acquire high-resolution UV echelle spectra with the \emph{Space Telescope
Imaging Spectrograph} on board the \emph{Hubble Space Telescope}
(\emph{HST/STIS}).

\section{Spectral line fits end elemental abundances}

As input for our spectrum synthesis, we used metal line-blanketed LTE model
atmospheres. 
Oscillator strengths were taken from
the Kurucz line list, as were damping constants for all metal
lines. Data for elements heavier than Zn were taken from literature. 
Only lines that have been observed experimentally were used, since we
required the most accurate wavelengths possible. 



\begin{figure}
\begin{center}
\includegraphics[width=1.00\textwidth]{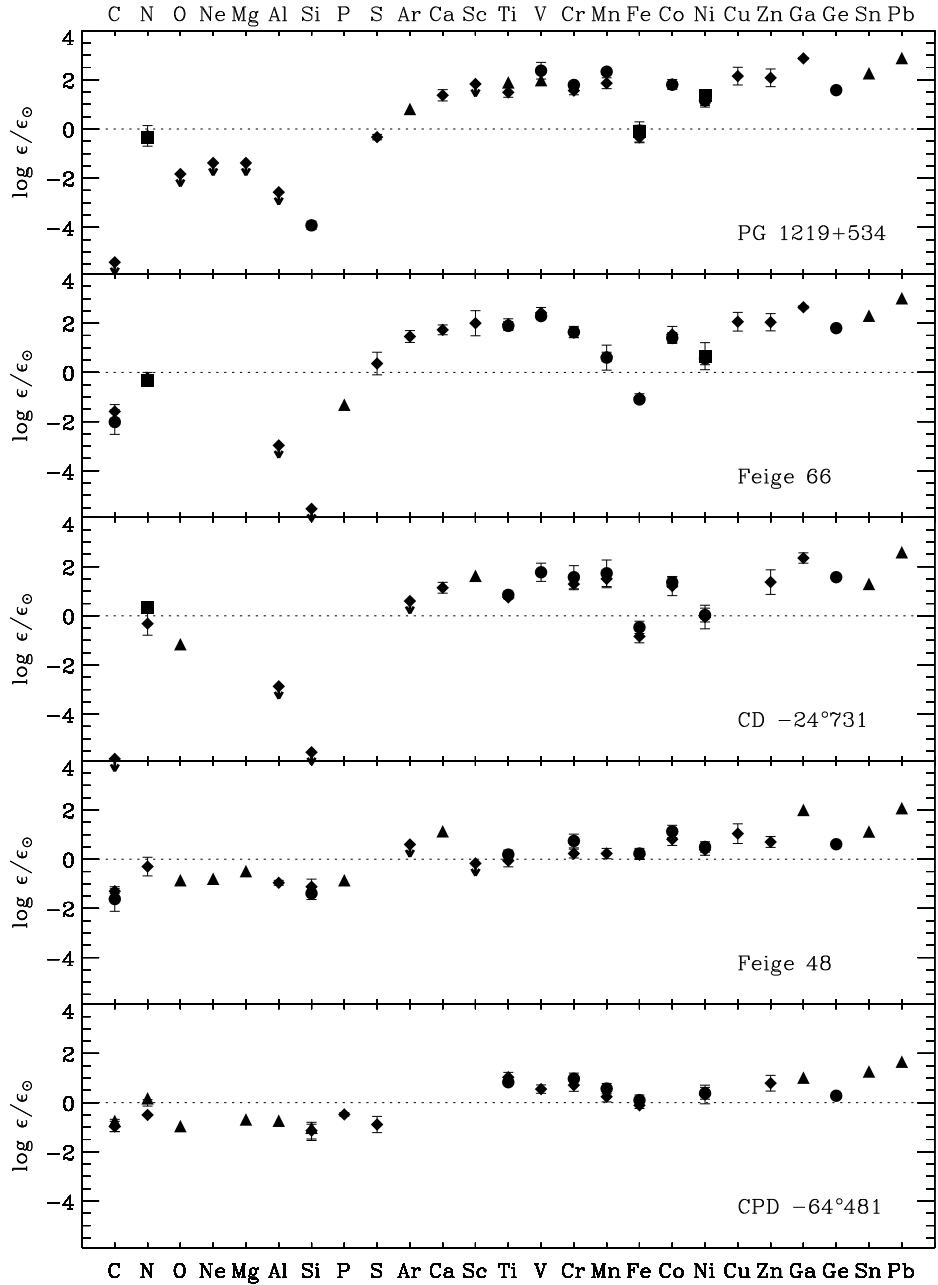}
\end{center}
\caption{Abundances measured for our five targets. Triangles
  represent values determined using singly ionised species, filled circles
  doubly ionised ones, diamonds
  triply ionised ones, and squares denote quadruply ionised
  species. Upper limits are marked by a downward arrow. 
  Note the generally excellent
agreement between different ionisation stages (from O'Toole \& Heber, 2006).}
\label{fig2}
\end{figure}

The results of the quantitative spectral analysis of our HST UV echelle 
spectra (O'Toole \& Heber, 2006) are plotted in Fig.~\ref{fig2}. 
Out of the five hot
sdB stars two are member of the short-period, pulsating
V361\,Hya class. 

Abundances of no less than 25 elements including the iron group and 
  even heavier elements such as tin and lead have been determined 
  (for details see O'Toole \& Heber, 2006). 
  
As has been found by many previous studies, carbon abundances range
from virtually none at all to slightly below the solar value while the N 
abundances are slightly below the solar
value. While Fe is found to be nearly solar (\pg, Feige~48, \cpd),
slightly depleted in \cd\ and subsolar in Feige~66 by a factor of ten, all
other elements of the iron group are enhanced by between 0.5 and 2.5 dex with
respect to solar values. The enhancements are large in Feige~66 and \pg, but
mild for the three others. The heavy metals Ga, Ge, Sn and Pb are all enriched 
with respect to the Sun in all stars, reaching as high as 2.9~dex for Ga in 
\pg\ or 2.75~dex for Pb in Feige~66. These peculiar abundance patterns are
probably caused by atomic diffusion processes in the atmosphere.

  We have compared a hot pulsator (\pg) with a
two non-pulsators with similar stellar parameters (Feige~66 and \cd) 
and a cooler pulsator
(Feige~48) with a similar non-pulsator (\cpd), and found no consistent
differences between the members of each pair. 

The heavy element abundance pattern of \cd\ comes close to that
observed for \pg\ except for its low Fe and Ni. Feige~66 has an even 
lower
Fe abundance, but its heavy metal abundance pattern does not match that of
\pg\ at all. In other words the abundance patterns of two non-variable stars 
of similar temperature and gravity are too dissimilar for a conclusive
comparison with a pulsator. 

This result
leads us to suspect
that there must be another, as yet unknown, discriminating factor between
pulsating and non-pulsating sdB stars.

\section{Lead Isotopes and the $^{207}$Pb/\pb\ ratio}

Measuring isotope ratios in stars is of key importance in
understanding the processes governing nucleosynthesis of the elements.
Unfortunately such measurements are difficult in stars since there are only a
handful of elements where the isotope splitting of spectral lines is
large enough to be measured. The isotopes of Pb have been
previously investigated in metal-poor halo stars 
as well as in a main-sequence B star in the SMC.

The lead isotope \pb\ is the terminal point of the decay of the
radioactive actinide sequence -- those elements that are formed only by the
$r$-process. Nucleosynthesis via the $s$-process also has a final peak
at \pb. This makes
lead one of the most important elements in nucleosynthesis
modeling.  

In metal-poor halo stars lead isotopic ratios have been measured from 
very few lines of neutral lead (Pb\,\textsc{i}). 
In hot stars, however, such as sdB stars
lead  is highly ionized and the resonance lines of triply ionized lead,
Pb\,\textsc{iv}, are expected to be the strongest.

Very high resolution UV spectra ($R=114\,000$, E140H grating of
\emph{HST/STIS}) of the sdB stars Feige~66 and \cpd\ covering the wavelength 
range
1160-1361\,\AA\ were retrieved from the MAST archive. 
The spectral line of interested in the Pb\,\textsc{iv} resonance
line at 1313.07\,\AA, which is
one of two Pb\,\textsc{iv} resonance lines; the other is at
1028.61\,\AA, outside the spectral window of HST.

\begin{figure}
\includegraphics[width=.5\textwidth]{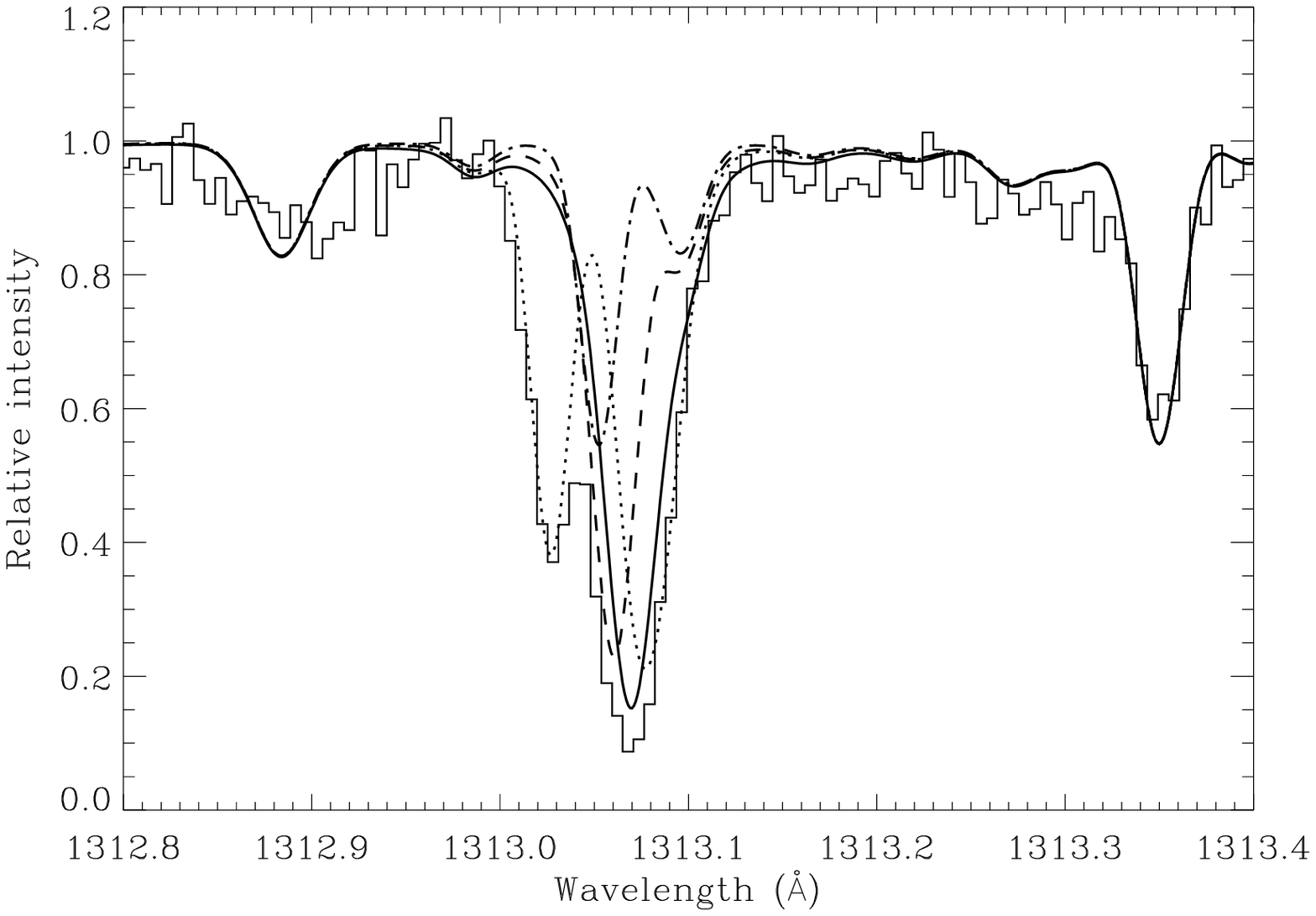}
\includegraphics[width=.5\textwidth]{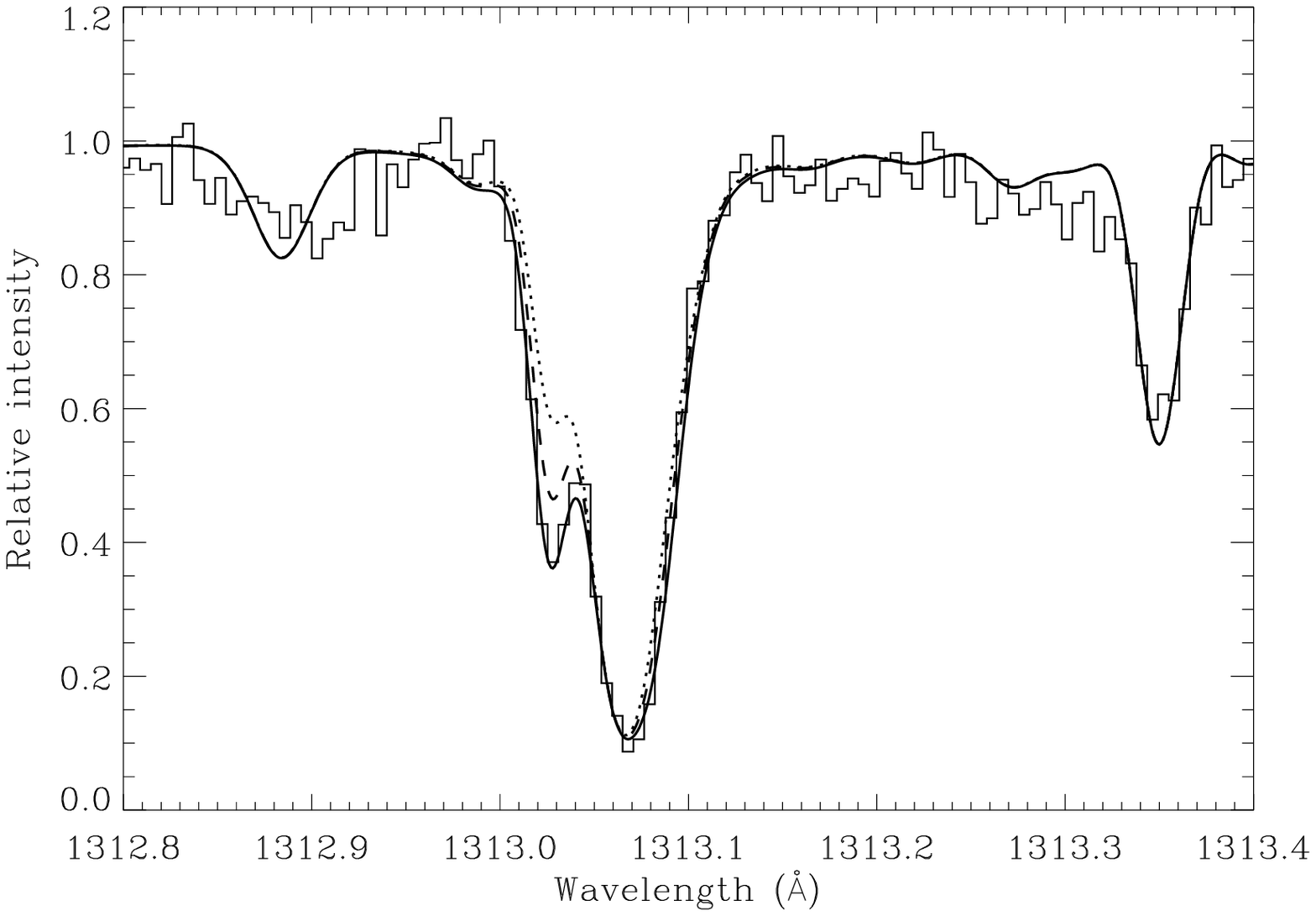}
\caption{
left hand side: The individual isotope contributions to the Pb\,\textsc{iv} resonance line in
\cpd\ are shown.
The observed spectrum is shown as a histogram, with the isotopes
shown as follows: $^{204}$Pb -- dash-dotted; $^{206}$Pb -- dashed; $^{207}$Pb
- dotted; $^{208}$Pb -- solid. As can be seen, the blue most component of the 
$^{207}$Pb line is easily resolved.
Right hand side: The Pb\,\textsc{iv} line profile of \cpd\ is shown. 
Over-plotted is our model calculated with the solar
system Pb isotope ratio, as well as 0.5 and 0.25 times the amount of
$^{207}$Pb. }
\label{fig3}
\end{figure}

The relative contribution of the $^{204}$Pb, $^{206}$Pb, $^{207}$Pb, and 
$^{207}$Pb to the line blend are shown in Fig.~\ref{fig3} (left hand panel)
assuming a solar system Pb isotope ratio. In the right hand panel 
of Fig.~\ref{fig3}
we depict the effect of reducing the $^{207}$Pb/$^{208}$Pb ratio to subsolar 
values.  
As can be seen the $^{207}$Pb/$^{208}$Pb ratio is solar. We cannot make any 
definitive
statements with regard to the other isotopes. In fact, our
observations are also consistent with \emph{no} $^{204}$Pb. We can
conclude though, that the solar system isotope mix is consistent with
our observations.





\end{document}